\documentclass[9pt,conference]{IEEEtran}
\IEEEoverridecommandlockouts
\usepackage[cmex10]{amsmath}
\usepackage{eqparbox}
\usepackage[usenames, dvipsnames]{color}
\usepackage{mathtools,amssymb,bm,mathabx,adjustbox}
\usepackage{amstext}
\usepackage{amssymb}
\usepackage{graphicx}
\usepackage{color}
\usepackage{booktabs}
\usepackage{longtable}
\usepackage{multicol}
\usepackage{algorithmicx}
\usepackage[ruled]{algorithm}
\usepackage{algpseudocode}
\usepackage{algpascal}
\usepackage{algc}
\usepackage{amsfonts}
\usepackage{dsfont}
\usepackage{array}
\usepackage[most]{tcolorbox}
\usepackage{cases}
\usepackage{pgfplots}
\usepackage{caption}
\usepackage{subfigmat}
\usepackage{xr-hyper}
\usepackage{hyperref}
\usepackage{amsthm}
\DeclareCaptionLabelSeparator{periodspace}{.~}
\usepackage{float}
\usepackage{cite}
\usepackage{epstopdf}
\theoremstyle{remark}

\newcommand\ASTART{\bigskip\noindent\begin{minipage}[b]{0.5\linewidth}}
	
	\newcommand\AENDSKIP{\end{minipage}\bigskip}
\newcommand\AEND{\end{minipage}}
\ifCLASSOPTIONcaptionsoff
\usepackage[nomarkers]{endfloat}
\let\MYoriglatexcaption\caption
\renewcommand{\caption}[2][\relax]{\MYoriglatexcaption[#2]{#2}}
\fi
\theoremstyle{plain}

\newtheorem{prop}{\textbf{Proposition}}

\theoremstyle{definition}

\newcommand*{\rom}[1]{\expandafter\@slowromancap\romannumeral #1@}

\newcommand{\RN}[1]{%
\textup{\uppercase\expandafter{\romannumeral#1}}%
}

\usepackage{standalone}
\graphicspath{{./Figures/}}
\usepackage{times}
\usepackage[T1]{fontenc}
\usepackage[english]{babel}
\usepackage{graphicx}
\usepackage{ amsfonts, dsfont,amssymb, graphicx, array, tabularx, booktabs,multicol,nccmath}
\usepackage{amsthm}
\usepackage{epsf,epsfig}
\usepackage{setspace}
\usepackage{subfigure}

\usepackage{hyperref}
\usepackage{multirow}
\usepackage{url}
\usepackage{color}
\usepackage[nolist,printonlyused]{acronym}      
\usepackage{comment}
\usepackage{cite}
\usepackage{bm}
\usepackage{tikz}
\usetikzlibrary{decorations.pathreplacing}

\usepackage{mathtools}

\usepackage{blindtext}

\usepackage{stfloats}

\newcommand{\gf}[1]{\textcolor{black}{{#1}}}

\newcommand{\mx}[1]{\mathbf{#1}}
\newcommand{\bs}[1]{\boldsymbol{#1}}

\definecolor{amber}{rgb}{1.0, 0.49, 0.0}
\definecolor{ao}{rgb}{0.0, 0.5, 0.0}

\def\R2#1{\textcolor{black}{#1}}
\def\R3#1{\textcolor{black}{#1}}

\renewcommand{\triangleq}{\mathbin{\setstackgap{S}{0pt}\stackMath\Shortstack{\smalltriangleup\\ =}}}
\usepackage[usestackEOL]{stackengine} 
\usepackage[utf8]{inputenc} 
\usepackage{url}
\usepackage{ifthen}
\usepackage{cite}

\usepackage{xcolor}
\def\BibTeX{{\rm B\kern-.05em{\sc i\kern-.025em b}\kern-.08em
    T\kern-.1667em\lower.7ex\hbox{E}\kern-.125emX}}
\begin{document}
\title{Near-Field ISAC in 6G: Addressing Phase Nonlinearity via Lifted Super-Resolution\\
\thanks{This work is supported in part by Digital Futures. G.Fodor was also supported by the Swedish Strategic Research (SSF) grant for the FUS21-0004 SAICOM project and the 6G-Multiband Wireless and Optical Signalling for Integrated Communications, Sensing and Localization (6G-MUSICAL) EU Horizon 2023 project, funded by the EU, Project ID: 101139176.}
}

\author{
    \IEEEauthorblockN{Sajad Daei \IEEEauthorrefmark{1}, \, Amirreza Zamani \IEEEauthorrefmark{1}, \, Saikat Chatterjee \IEEEauthorrefmark{1}, \, Mikael Skoglund \IEEEauthorrefmark{1}, \, Gabor Fodor \IEEEauthorrefmark{1}\IEEEauthorrefmark{2}} 
    
    \IEEEauthorblockA{\IEEEauthorrefmark{1} Digital Futures Centre and School of Elect. Engg. $\&$ Comp. Sc., KTH Royal Institute of Technology, Sweden}  
    \IEEEauthorblockA{\IEEEauthorrefmark{2} Ericsson Research, Stockholm, Sweden} 
    \IEEEauthorblockA{sajado@kth.se, \,  amizam@kth.se, \,sach@kth.se, \, skoglund@kth.se, \, gaborf@kth.se}
}

\maketitle


\begin{abstract}
\gf{Integrated sensing and communications} (ISAC) is a promising component of 6G networks, 
\gf{fusing communication and radar technologies to facilitate new services.}
Additionally, the use of extremely large-scale antenna arrays (ELAA) at the ISAC common receiver not only facilitates terahertz-rate communication links but also significantly enhances the accuracy of target detection in radar applications. In practical scenarios, communication scatterers and radar targets often reside in close proximity to the ISAC receiver. This, combined with the use of ELAA, fundamentally alters the electromagnetic characteristics of wireless and radar channels, shifting from far-field planar-wave propagation to near-field spherical wave propagation. Under the far-field planar-wave model, the phase of the array response vector varies linearly with the antenna index. In contrast, in the near-field spherical wave model, this phase relationship becomes nonlinear. This shift presents a fundamental challenge: the widely-used Fourier analysis can no longer be directly applied for target detection and communication channel estimation at the ISAC common receiver. In this work, we propose a feasible solution to address this fundamental issue. Specifically, we demonstrate that there exists a high-dimensional space in which the phase nonlinearity can be expressed as linear. Leveraging this insight, we develop a lifted super-resolution framework  that simultaneously performs communication channel estimation and extracts target parameters with high precision.
\end{abstract}

\section{Introduction}
\gf{Integrated sensing and communication (ISAC) has emerged as a promising} technology in the evolution of next-generation wireless systems, enabling the simultaneous execution of radar sensing and communication tasks within a unified framework. As modern applications such as autonomous vehicles, smart homes, and industrial automation continue to advance, the ability to seamlessly integrate these functionalities becomes increasingly critical. An ISAC receiver is tasked not only with decoding messages from communication users but also with detecting and tracking radar targets in its vicinity. This dual capability necessitates the development of sophisticated techniques to enhance the performance of both sensing and communication functions 
\cite{lu2024tutorial,huawei2026g,liu2024near,zhang2023near,lu2022near,li2024near}.
One promising approach for improving ISAC performance is the employment of extremely large-scale antenna arrays (ELAA) at the receiver. ELAA can significantly enhance the accuracy of sensing and detecting nearby objects, as well as increase communication capacity by leveraging a large number of antennas. For instance, an ELAA deployed on a building facade can effectively sense pedestrians and vehicles while simultaneously supporting communication at terabit-per-second (Tbps) data rates.

\begin{figure}
    \centering
    \includegraphics[scale=0.2]{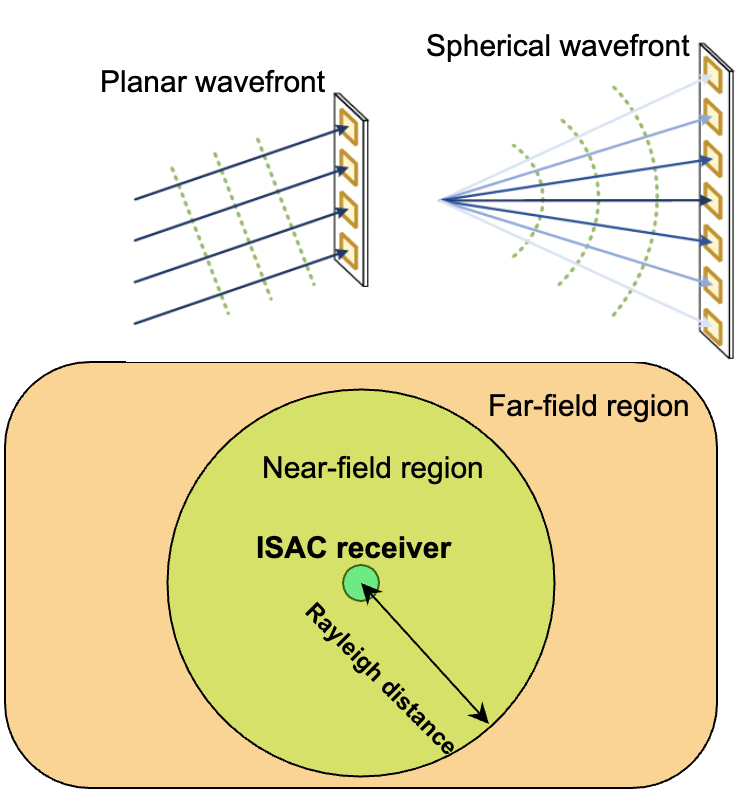}
    \caption{Top image: Planar wavefront in far-field versus spherical wavefront in near-field. Bottom image: This figure shows the near-field region around an ISAC receiver which is known as Rayleigh distance. }
    \label{fig:rayleigh_distance}
\end{figure}

Traditionally, ISAC systems have operated under the assumption that radar targets and communication users are positioned in the far-field range, defined as distances greater than the Rayleigh distance, $\frac{2D^2 f_c}{c}$, where $c=3\times 10^8$ m/s is the speed of light, $f_c$ is the carrier frequency, and $D$ is the array aperture. For a typical array aperture of $D=4$ meters, often deployed in building facades, and a carrier frequency of $f_c=28$ GHz, this Rayleigh distance is approximately 2987 meters. However, the assumption that the communication scatterers and radar targets are located at such distances from the ISAC receiver (e.g., the building in our example) is impractical in many real-world scenarios. This limitation motivates the consideration of near-field ISAC, where the echo signals from targets and communication scatterers are modeled as spherical waves, in contrast to the planar wavefronts assumed in far-field conditions (see Figure \ref{fig:rayleigh_distance}).

Near-field ISAC presents several notable advantages over its far-field counterpart. 
Firstly, the limited range and focused signal propagation of the system significantly enhance security by minimizing the risk of eavesdropping and ensuring that sensitive data remains confined to its intended area of operation. In such scenarios, statistical privacy designs, as explored in \cite{amir1, amir2}, can be particularly effective. This localized approach also improves privacy by minimizing the chance of signal interception by unintended receivers. Such attributes make near-field ISAC especially suitable for applications such as secure device-to-device data transfers, and personal health monitoring, where proximity adds an extra layer of security and privacy.

Furthermore, the spherical wavefronts characteristic of near-field interactions provide superior spatial resolution for sensing and enable higher data rates for communication. This is achieved by allowing multiple data streams to be transmitted simultaneously across distinct spatial paths. Near-field ISAC systems also benefit from lower power consumption due to shorter transmission distances and more efficient signal processing, making them particularly advantageous for battery-powered or portable devices. Additionally, the localized nature of near-field interactions helps to mitigate interference from external sources, reducing the likelihood of signal crosstalk and resulting in clearer, more reliable communication channels.

Despite these benefits, the transition from far-field to near-field ISAC introduces unique challenges, particularly concerning the phase characteristics of the channel's steering vectors. Unlike in the far field, where the phase of the steering vectors exhibits linear behavior, near-field conditions introduce nonlinear phase variations that complicate signal processing and system design. Effectively addressing these phase nonlinearities is essential for realizing the full potential of near-field ISAC. 

In this paper, we introduce \textit{lifted} super-resolution, a novel approach to manage phase nonlinearities in near-field ISAC channels. Our method involves a technique that effectively transforms the nonlinear problem into a linear problem with higher-dimensional space, where linearization becomes achievable. We then apply a super-resolution technique to promote the continuous angular sparsity that arises from the continuous nature of the angles corresponding to radar targets and communication scatterers. Finally, we estimate the distances of these targets and scatterers using a least squares optimization method. Through rigorous analysis and simulations, we demonstrate that our approach not only mitigates the impact of phase nonlinearities but also enhances the overall sensing and communication performance of near-field ISAC, even with limited resources.

\textit{Notation}.
$\mx{I}_N$ represents the identity matrix of size $N$. The column stacking vector operator, denoted as $\textbf{vec}$, transforms a matrix $\mx{X}\in\mathbb{C}^{M\times N}$ into its vectorized version $\mx{x}\triangleq \textbf{vec}(\mx{X})\in\mathbb{C}^{M N\times 1}$. ${\rm vec}^{-1}(\cdot)$ reshapes a vector $\mx{x}$ into a matrix $\mx{X}$ by stacking the columns vertically. The ceil operator ${\rm ceil}(x)$ is a mathematical functions that rounds a given number $x$ up to the smallest integer greater than or equal to that number. The growth rate of a function $f(r)$ in terms of another function $g(r)$ is expressed as $f(r)=\textit{o}(g(r))$ which implies that $\lim_{r\rightarrow \infty} \frac{f(r)}{g(r)}=0$.
The $(k,l)$-th element of the matrix $\mx{A}$ is denoted as $[\mx{A}]_{k,l}$. Also, the $k$-th element of a vector $\mx{x}\in\mathbb{C}^N$ is shown by $x[k]$. The maximum singular value of a matrix $\mx{X}$ is shown by $\sigma_{\max}(\mx{X})$. $1_{\mathcal{E}}$ stands for the indicator function which is one when the event $\mathcal{E}$ holds and zero elsewhere. The Bessel function of order $n$ is defined as $J_n(x) \triangleq \frac{1}{\pi} \int_0^\pi \cos(nt - x \sin(t)) \, {\rm d} t $. 
\begin{figure}
    \centering
    \includegraphics[scale=.3]{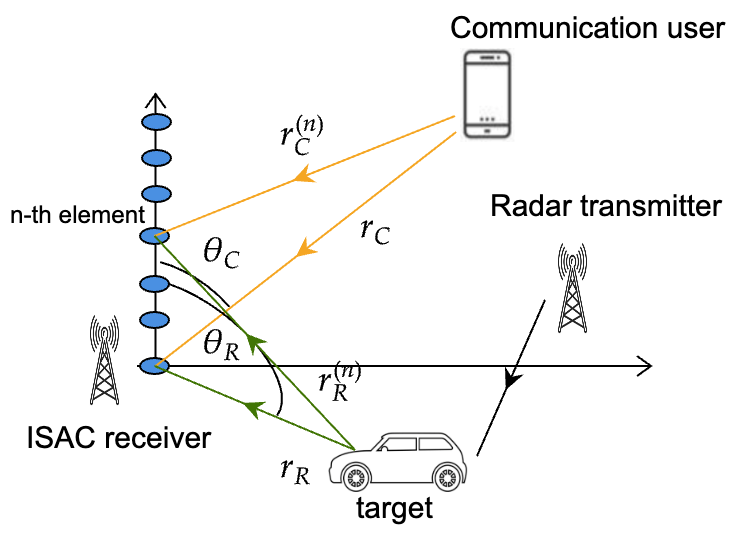}
    \caption{This figure illustrates the ISAC model under consideration, featuring an ISAC receiver equipped with ELLA, a communication user, and a target. The echo from the target and the message from the user simultaneously reach the ISAC receiver.}
    \label{fig:model}
\end{figure}
\section{System model}
\label{sec:system}
We consider a near-field ISAC system with $s_R$ targets, one user, and a common ISAC receiver equipped with uniform linear array. There are $s_C$ scatterers between the user and the ISAC receiver known as communication scatterers. A specific and simple example of this scenario, with $s_C = s_R = 1$, is depicted in Figure \ref{fig:model}. The number of antennas employed at the ISAC receiver is denoted by $N_r$, each spaced by a distance $d$, while the communication user is equipped with a single antenna. The communication channel between the user and the ISAC receiver is stated as
\begin{align}\label{eq:comm_channel}
 \mx{h}_{\rm C}=\sum_{i=1}^{s_c} \alpha^i_{\rm C}\mx{a}(\theta^i_{C}, r^i_{C})\in\mathbb{C}^{N_r\times 1}
\end{align}
while the radar sensing channel is stated as:
\begin{align}\label{eq:rad_channel}
    \mx{h}_{\rm R}= \sum_{i=1}^{s_R}\alpha^i_{\rm R}\mx{a}(\theta^i_{R}, r^i_{R}) \in\mathbb{C}^{N_r\times 1} 
\end{align}
Here, $\alpha^i_C$, $\alpha^i_R$ $\theta^i_C$, $\theta^i_R$ are the channel gains and angles for communication and radar, respectively corresponding to $i$-th path, and 
\begin{align}\label{eq:steering_vec}
 \mx{a}(\theta,r)\triangleq \begin{bmatrix}
     1,{\rm e}^{-j\frac{2\pi}{\lambda}(r^{(2)}-r)},..., {\rm e}^{-j\frac{2\pi}{\lambda}(r^{(N_r)}-r)  }
 \end{bmatrix}^T  \in\mathbb{C}^{N_r\times 1}
\end{align}
is the array response vector which depends on both angle and distance. For simplicity and notational convenience, we focus on a single line-of-sight (LoS) channel ($s_C = 1$) and one target ($s_R = 1$). The proposed approach can be extended to more complex scenarios. The distance of the target relative to the reference antenna is connected to the distance relative to $n$-th antenna as follows:
\begin{align}\label{eq:distance_formula}
    r_R^{(n)}=\sqrt{r_R^2+n^2 d^2-2 r_R n d\cos(\theta_R)}, n=0,..., N_r-1
\end{align}
where $r_R\triangleq r_R^{(1)}$ and $\theta_R\triangleq \theta_R^{(1)}$ are the distance and angle of the target with respect to the reference antenna. Similarly, $r_C\triangleq r_C^{(1)}$ and $\theta_C\triangleq \theta_C^{(1)}$ are the distance and angle of the user to the reference antenna. 
After collecting $m$ samples at the ISAC receiver, the resulting received signal can be expressed as \cite{wang2023near,zhang2023near,sayyari2020blind,daei2024timely}:
\begin{align}\label{eq:measurements}
 \mx{y}=\mx{A}_{C} \mx{h}_C+\mx{A}_{R} \mx{h}_R +\bs{\epsilon}\in\mathbb{C}^{m\times 1}
\end{align}
where $\mx{A}_{C}\in\mathbb{C}^{m\times N_r}$ is the known pilot matrix for estimating the communication channel and $\mx{A}_{R} \in\mathbb{C}^{m\times N_r}$ is the known radar matrix composed of radar waveform transmitted by radar transmitter. $\bs{\epsilon}$ is the additive noise. The goal is to estimate the continuous-valued distance-angle pair for both radar target and communication user.

\section{Proposed method}
The equation \eqref{eq:steering_vec} shows that the phase of the steering vector $\mx{a}(\theta,r)$ depends nonlinearly on two continuous-valued variables: the angle $\theta$ and the distance $r$. This nonlinearity poses significant challenges in applying the well-established linear Fourier analysis for recovering the channel and targets, as discussed in several works, including \cite{cui2022channel,zhang2023near,lu2014overview,liu2024near,wang2023near}. Another issue is that the angle-distance pair are continuous-valued parameters and can take any arbitrary values and can not be assumed to lie on predefined domain of grids \cite{zhang2023near}. In this work, we first employ a lifted transformation to handle the phase nonlinearity and then propose an optimization technique to recover the continuous-valued angle and distance parameters. To do so, we first use the following Taylor expansion:
\begin{align}
    \sqrt{1+x}=1+\frac{x}{2}-\frac{x^2}{8}+...
\end{align}
to write
\begin{align}\label{eq:rll1}
  &\sqrt{r^2+n^2 d^2-2 r n d\cos(\theta)}  \approx\nonumber\\
  &r-n d \cos(\theta)+\frac{n^2d^2}{2 r}-\frac{r(\frac{n^2 d^2}{r^2}-2\frac{n d \cos(\theta)}{r})^2}{8}\approx\nonumber\\
  &r(1-\frac{n d \cos(\theta)}{r}+\frac{n^2 d^2}{2 r^2}\sin^2(\theta))+\textit{o}(\frac{1}{r}).
\end{align}

Here, we only use the first three terms, which correspond to the Fresnel distance \cite{liu2023near}. However, the proposed approach can be extended to include more terms, albeit at the cost of increased complexity. By using some simple trigonometric properties and the relation \eqref{eq:rll1}, the $n$-th element of the steering vector can be written as
\begin{align}
    \mx{a}(\theta, r)[n]\approx {\rm e}^{-j\frac{2\pi}{\lambda}\frac{n^2d^2}{4 r}}{\rm e}^{-j\frac{2\pi}{\lambda}(-n d \cos(\theta)-\frac{n^2 d^2}{4 r}\cos(2\theta))}.
\end{align}
We observe that the phase depends nonlinearly on both $\theta$ and $r$. To address the nonlinearity with respect to $\theta$, we employ the Jacobi-Anger expansion:
\begin{align}\label{eq:jacobi}
    {\rm e}^{j {z}\cos(\theta)}=\sum_{l=-\infty}^{\infty} j^{l}J_l(z){\rm e}^{l \theta}.
\end{align}

This expansion effectively transforms the problem into a higher-dimensional space where linearization becomes feasible. As a result, it facilitates the application of linear processing techniques, such as super-resolution methods, to accurately recover the continuous-valued angles. After applying the Jacobi-Anger expansion, the steering vector becomes in the form of

\begin{align}\label{eq:steer_approx}
   \scalebox{.8}{$\mx{a}(\theta, r)[n]\approx {\rm e}^{-j\frac{2\pi}{\lambda}\frac{n^2d^2}{4 r}}\sum_{l=-\infty}^{\infty}\sum_{q=-\infty}^{\infty} j^{l+q} J_l(\frac{2\pi n d }{\lambda}) J_q(\frac{2\pi}{\lambda}\frac{n^2d^2}{4 r}) {\rm e}^{j (l+2 q)\theta }$}.
\end{align}
The formulation \eqref{eq:steer_approx}
provides a way to resolve the nonlinearlity at the cost of higher complexity. For a fixed $x$, the Bessel function $J_n(x)$ exhibits a decaying behaviour in terms of the order $n$. For large $n$, the bessel function can be approximated as $J_n(x)\approx \frac{1}{\sqrt{2 \pi n}}(\frac{ e x}{2 n})^n$ where $e$ refers to Euler's number. When $n$ is large, $(\frac{ e x}{2 n})^n$ approaches zero when $ \frac{ e x}{2 n}< 1$. This suggests that Bessel function $J_n(x)$ will become very small for orders $n$ greater than $n_0\approx {\rm ceil}(\frac{ e x}{2})$. By having this in mind, the steering vector in \eqref{eq:steer_approx} can be further approximated as follows:
\begin{align}
    \scalebox{.8}{$\mx{a}(\theta, r)[n]\approx {\rm e}^{-j\frac{2\pi}{\lambda}\frac{n^2d^2}{4 r}}\sum_{l=-I_1}^{I_1}\sum_{q=-I_2}^{I_2} j^{l+q} J_l(\frac{2\pi n d }{\lambda}) J_q(\frac{2\pi}{\lambda}\frac{n^2d^2}{4 r}) {\rm e}^{j (l+2 q)\theta } $}
\end{align}
where $I_1={\rm ceil}(\frac{e\pi (N_r-1) d}{\lambda})$ and $I_2={\rm ceil}(\frac{e\pi (N_r-1)^2 d^2}{4 r\lambda})$. The above formulation can be stated as the inner product of two matrices as

\begin{align}
 &\mx{a}(\theta, r)[n] \approx \langle \bs{\Gamma}_n(r), \mx{V}(\theta) \rangle =\langle {{\rm vec}(\bs{\Gamma}_n(r))}, {{\rm vec}(\mx{V}(\theta))} \rangle\nonumber\\
 &\triangleq \langle \bs{\gamma}_n(r),\mx{v}(\theta) \rangle, n=0,..., N_r-1
\end{align}
where 
\begin{align}
    \bs{\Gamma}_n(r)[l,q]\triangleq {\rm e}^{-j\frac{2\pi}{\lambda}\frac{n^2d^2}{4 r}}j^{l+q} J_l(\frac{2\pi n d }{\lambda}) J_q(\frac{2\pi}{\lambda}\frac{n^2d^2}{4 r}) ,
\end{align}

\begin{align}
  \mx{V}(\theta)[l,q]\triangleq  {\rm e}^{j (l+2 q)\theta } 
\end{align}
 and $\mx{v}(\theta)\triangleq  \mx{V}(\theta), \bs{\gamma}_n(r)\triangleq {\rm vec}(\bs{\Gamma}_n(r))$. By collecting all samples $n=0,..., N_r-1$ and defining $I_1^\prime\triangleq 2 I_1+1, I_2^\prime \triangleq 2I_2+1$, the steering vector can be expressed as
 \begin{align}\label{eq:steering_vec_new}
      \mx{a}(\theta, r)=\mx{C}(r) \mx{v}(\theta)\in\mathbb{C}^{N_r\times 1},
 \end{align}
 where $\mx{C}(r)\triangleq [\gamma_{0}(r), ...,  \gamma_{N_r-1}(r)]^T\in\mathbb{C}^{N_r\times I_1^\prime I_2^\prime }$ is called the distance matrix depending only on the distance and $\mx{v}(\theta)$ is called far-field steering vector depending only on $\theta$. 
\eqref{eq:steering_vec_new} shows that the steering vector in near-field can stated as the multiplication of a distance-dependent matrix and a Vandermonde vector showing the steering vector of far-field. We refer to this transformation as the lifted transformation. In the latter expression, both distance and angles are continuous-valued parameters. Forcing the angles to be on a predefined grid can substantially degrade the recovery performance leading to the well-known basis mismatch concept \cite{zhang2023near,chi2020harnessing, daei2023blind}. To promote the angular continuous sparsity, we first propose an optimization problem to estimate the continuous-valued angle parameter. Merging the expression \eqref{eq:steering_vec_new} with \eqref{eq:comm_channel} and \eqref{eq:rad_channel}, the communication and radar channels can be described as
\begin{eqnarray}
\begin{array}{c}
\mx{h}_C=\displaystyle\sum_{i=1}^{s_C} \alpha^i_C \mx{C}_C(r_C^i)\mx{v}(\theta_C^i)\in\mathbb{C}^{N_r\times 1}, \\
\mx{h}_R=\displaystyle\sum_{i=1}^{s_R}\alpha^i_R \mx{C}_{R}(r_R^i)\mx{v}(\theta^i_{R})\in\mathbb{C}^{N_r\times 1},
\end{array}
\end{eqnarray}
where the pairs $(\theta^i_C,r^i_C)$ and $(\theta^i_R,r^i_R)$ refers to the angle-distance of $i$-th communication scatterer and radar target, respectively. Here, $\mx{C}_C(r^i_C)\in \mathbb{C}^{N_r\times I_{1,C}^\prime {I^{\prime}_{2,i,C}}}$ and $\mx{C}_R(r^i_R)\in \mathbb{C}^{N_r\times I_{1,R}^\prime {I^{\prime}_{2,i,R}}} $ are the communication and radar distance matrices corresponding to $i$-th scatterer and target, respectively. To promote the angular continuous sparsity, the following optimization is proposed:
\begin{align}\label{prob.main}
&\min_{\mx{z}_C,\mx{z}_R} \|\mx{z}_C\|_{\mathcal{A}_C}+ \|\mx{z}_R\|_{\mathcal{A}_R}~~{\rm s.t.}~~ \|\mx{y}-\mx{A}_C \mx{z}_C-\mx{A}_R \mx{z}_R \|_2\le \eta
\end{align}
where $\|\mx{z}_C\|_{\mathcal{A}_C}$ and $\|\mx{z}_R\|_{\mathcal{A}_R}$ are called atomic norms \cite{chandrasekaran2012convex,candes2014towards,tang2013compressed,valiulahi2019two}, that encourage continuous angular sparsity for communication and radar defined respectively as
\begin{align}
 \scalebox{.8}{$ \|\mx{z}_C\|_{\mathcal{A}_C}\triangleq \inf \{ \sum_i |\alpha_i| :~\mx{z}_C=\sum_i \alpha_i \mx{C}_C(r_i)v(\theta), \theta\in (0,\pi), r_i\in(r_{\min,C},r_{\max,C}) \} $} 
\end{align}
and 
\begin{align}
  \scalebox{.8}{$\|\mx{z}_R\|_{\mathcal{A}_R}\triangleq \inf \{ \sum_i |\alpha_i| :~\mx{z}_R=\sum_i \alpha_i \mx{C}_R(r_i)v(\theta), \theta\in (0,\pi), r_i\in(r_{\min,R},r_{\max,R}) \}  $}
\end{align}
For simplicity, we consider one target and one user with LoS channel, i.e., $s_C=s_R=1$. Here, $r_{\min,R},r_{\max,R}$ represent the minimum and maximum possible estimates for radar targets. Similarly, $r_{\min,C},r_{\max,C}$ represent the corresponding values for communication user. $\eta$ is an upper-bound for $\|\bs{\epsilon}\|_2$
The dual of the latter optimization  problem can be obtained as follows (see e.g. \cite{daei2024timely,daei2023blind,daei2023blind_wiopt,safari2021off,maskan2023demixing,seidi2022novel}):
\begin{align}\label{prob.dual}
&\max_{\mx{q}} {\rm Re} \langle \mx{y},\mx{q} \rangle -\eta \|\mx{q}\|_2\nonumber\\
&~{\rm s.t.}~ \|\mx{A}_C^H \mx{q}\|_{\mathcal{A}_C}^{\text{d}}\le 1, \|\mx{A}_R^H \mx{q}\|_{\mathcal{A}_R}^{\text{d}}\le 1,
\end{align}
where $\|\mx{x}\|_{\mathcal{A}_C}^{\text{d}}\triangleq\sup_{\substack{r\in[r_{\min,C},r_{\max,C}]\\\theta\in(0,\pi)}} \langle \mx{x}, \mx{C}_C(r)\mx{v}(\theta)\rangle  $ is defined as the dual atomic function for communication part. A similar definition applies to the radar part. Since both distances and angles are continuous-valued, the problem in \eqref{prob.dual} involves an infinite number of constraints, making it intractable. To address this, we propose a tractable semidefinite programming (SDP) approach that converts the infinite constraints into a set of linear matrix inequalities without compromising accuracy. This approach is inspired by the theory of positive trigonometric polynomial initially proposed in \cite{dumitrescu2017positive,candes2014towards}.
\begin{prop}\label{prop.sdp}
The optimization problem \eqref{prob.dual} can be relaxed to the following SDP optimization problem:
\begin{align}\label{prob.sdp}
&\max_{\mx{Q}_C,\mx{Q}_R,\mx{q}} {\rm Re} \langle \mx{y},\mx{q} \rangle-\eta \|\mx{q}\|_2\nonumber\\
&{\rm s.t.}~ \scalebox{.8}{$\begin{bmatrix}
\mx{Q}_C&  \mx{C}^{H}_C(r_{\max,C})\mx{A}_C^H \mx{q}\\
\mx{q}^H \mx{A}_C\mx{C}_C(r_{\max,C})&1
\end{bmatrix}\succeq \mx{0}$}\nonumber\\
&\scalebox{.8}{$\begin{bmatrix}
\mx{Q}_R& \mx{C}^{H}_R(r_{\max,R})\mx{A}_R^H \mx{q}\\
\mx{q}^H \mx{A}_R\mx{C}_R(r_{\max,R}) &1
\end{bmatrix}\succeq \mx{0} $} \nonumber\\
&\langle \bs{\Theta}_{n_1},\mx{Q}_C\rangle=1_{n_1=0},\langle \bs{\Theta}_{n_2},\mx{Q}_R\rangle=1_{n_2=0},
\end{align}
where $\bs{\Theta}_n$ is the elementary Toeplitz matrix with ones along its $n$-th diagonal and zero elsewhere. $\mx{Q}_C$ is of size $I^{\prime}_{1,C}I^{\prime}_{2,C}\times I^{\prime}_{1,C}I^{\prime}_{2,C}$ where $I^{\prime}_{2,C}\triangleq I^{\prime}_{2,1,C}$.
\end{prop}
After solving the latter SDP problem, we can obtain the estimates for angles. Define the radar and communication dual polynomial respectively as follows:
\begin{align}
    &f_R(\theta)\triangleq \sup_{r\in[r_{\min, R},r_{\max, R}]} \langle \mx{A}_R^H \mx{q}, \mx{C}_R(r)\mx{v}(\theta)\rangle\label{eq:radar_dualpoly}\\
 &f_C(\theta)\triangleq \sup_{r\in[r_{\min, C},r_{\max, C}]} \langle \mx{A}_C^H \mx{q}, \mx{C}_C(r)\mx{v}(\theta)\rangle\label{eq:comm_dual_poly}
\end{align}
Based on the results provided in \cite{daei2024timely,daei2023blind,sayyari2020blind,razavikia2023off,daei2023blind_wiopt}, it can be shown that the angle estimates for target and communication scatterer (which in our example is the communication user) can be obtained via finding angles that maximizes $|f_R(\theta)|$ and $|f_C(\theta)|$, respectively (see e.g. Figure \ref{fig:dualpolynomial} that shows the dual polynomial functions for radar and communications).
After finding the angle estimates for targets and communication scatterer, we replace the angle estimates into \eqref{eq:measurements}. Then, we find the distance estimates by iterating the following alternative optimization problems:
\begin{align}\label{eq:distance_estimate}
&\scalebox{.9}{$
\{\widehat{r}_C, \widehat{r}_R\}=\mathop{\arg\min}_{r_C,r_R} \|\mx{y}-\widehat{\alpha}_C\mx{A}_C \mx{a}(\widehat{\theta}_C, r_C)-\widehat{\alpha}_R\mx{A}_R \mx{a}(\widehat{\theta}_R, r_R)\|_2 . $}\nonumber\\
 &\scalebox{.9}{$\{\widehat{\alpha}_C,\widehat{\alpha}_R\}=\mathop{\arg\min}_{\alpha_C,\alpha_R} \|\mx{y}-\alpha_C\mx{A}_C \mx{a}(\widehat{\theta}_C, \widehat{r}_C)-\alpha_R\mx{A}_R \mx{a}(\widehat{\theta}_R, \widehat{r}_R)\|_2$}.
\end{align}
After a few iterations, the above method converges to the estimates of channel gains and distances. 
\section{Numerical Results}
In this section, we evaluate the performance of our method through a numerical experiment. We consider a target located at $\theta_R=\frac{\pi}{6}$ and $r_R=4$, and a user positioned at $\theta_C=\frac{\pi}{3}$ and $r_C=2$. The experiment uses $m=10$ measurements and $N_r=20$ antennas, with a carrier frequency set to $f_c=100$ GHz. The known pilot matrix for radar and communications are chosen randomly from i.i.d. Gaussian distribution.

\begin{figure}[t]
    \centering
    \includegraphics[scale=.4]{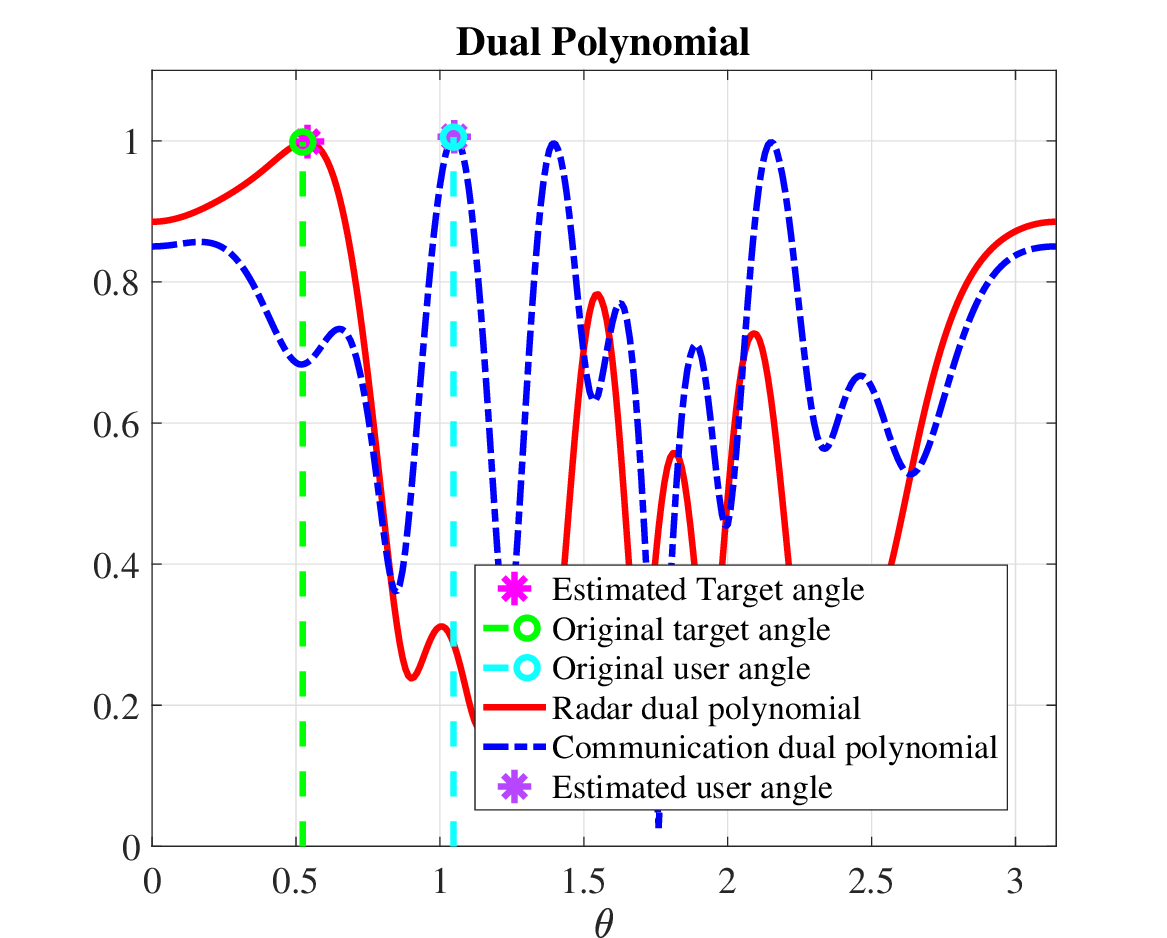}
    \caption{Dual polynomial plots for radar (red) and communications (blue). The original angles for the target and user are indicated by circles, while the estimated angles, corresponding to the maxima, are marked by asterisks.}
    \label{fig:dualpolynomial}
\end{figure}

We first solve the proposed SDP optimization \eqref{prob.sdp} to find the dual vector-valued variable $\mx{q}\in\mathbb{C}^{m\times 1}$. Then, we form the radar and communications dual polynomials shown by $|f_R(\theta)|$ and $|f_C(\theta)|$ defined in \eqref{eq:radar_dualpoly} and \eqref{eq:comm_dual_poly}.
Additionally, the distance estimates obtained using \eqref{eq:distance_estimate} are $\widehat{r}_C=1.9$ and $\widehat{r}_R=4.05$, which coincide with the ground-truth values.
Figure \ref{fig:dualpolynomial} displays these dual polynomial plots for radar and communications along with the ground-truth angles. The estimated angles for the target and user are obtained by identifying the angles that maximize $|f_R(\theta)|$ and $|f_C(\theta)|$. As it turns out from Figure \ref{fig:dualpolynomial}, the estimated angles perfectly match the ground-truth angles. Moreover, based on \eqref{eq:distance_estimate}, we find distance estimates which are obtained as $\widehat{r}_C=1.9$ and $\widehat{r}_R=4.05$ in this experiment which almost coincide with the ground-truth values of distance parameters. This also shows the capability of the proposed method in resolving the distances corresponding to the target and user.
It is important to note that the computational complexity of the SDP problem depends on the parameters $I_1={\rm ceil}(\frac{e\pi (N_r-1) d}{\lambda})$ and $I_2={\rm ceil}(\frac{e\pi (N_r-1)^2 d^2}{4 r\lambda})$ which in turn depends on the frequency, $N_r$ and the distance parameters for target and communication users. For example, if the targets are closer to the ISAC receiver or the carrier frequency is higher, the proposed strategy leads to increased computational complexity.

\section{Conclusions}
ISAC is positioned to become a critical component of 6G networks. ISAC systems equipped with extremely large antenna arrays offer not only enhanced communication capacity but also significantly improved target detection resolution. The proximity of targets and communication scatterers to the ISAC receiver in the near-field introduces a paradigm shift in the electromagnetic channel, transitioning from far-field planar wavefronts to near-field spherical wavefronts. In this paper, we proposed a novel approach for near-field ISAC receivers that enables simultaneous communication channel estimation and sensing. We demonstrated that the trigonometric nonlinearity present in the phase of near-field channels can be linearized through a lifted transformation, albeit with increased computational complexity. A super-resolution technique was proposed to recover the continuous-valued angles from a limited number of measurements at the ISAC receiver without requiring discretization of the angular domain. Given the estimated angles, we then employed an alternative optimization technique to recover the distances and channel gains. Numerical results validated the effectiveness of the proposed method in estimating angles and distances, highlighting its potential for practical near-field ISAC applications in 6G systems.

\section*{Appendix: Proof of Proposition \ref{prop.sdp}}\label{proof.sdp} 
First, we proceed with the dual atomic norm related to the communication term. The radar term follows the same strategy.
\begin{align}
\|\mx{A}_C^H \mx{q}\|_{\mathcal{A}_C}^{\rm d}\triangleq\sup_{r,\theta} \langle \mx{A}_C^H \mx{q}, \mx{C}_C(r)\mx{v}(\theta)\rangle    
\end{align}
For each distance  parameter $r$, we must have
\begin{align}\label{eq:condition1}
    \sup_{\theta} \langle \mx{C}_C^H(r)\mx{A}_C^H \mx{q}, \mx{v}(\theta)\rangle  \le 1
\end{align}
According to the theory of positive trigonometric polynomials \cite{dumitrescu2017positive} and Schur complement lemma \cite{horn2012matrix} for a non-negative matrix , \eqref{eq:condition1} is equivalent to the following linear matrix inequalities for each $r$:
\begin{align}\label{eq:condition2}
&\mx{Q}_C-\mx{C}_C^H(r)\mx{A}_C^H \mx{q} \mx{q}^H \mx{A}_C \mx{C}_C(r)\succeq \mx{0}, ~~\langle \bs{\Theta},\mx{Q}_C\rangle =1_{n_1=0}
\end{align}
Since
\begin{align}
   \lambda_{\min}(\mx{Q}_C)\ge \mx{q}^H \mx{A}_C\mx{C}_C(r_{\max,C}) \mx{C}_C^H(r_{\max,C})\mx{A}_C^H \mx{q}
\end{align}
is a sufficient condition to satisfy
\begin{align}
  \lambda_{\min}(\mx{Q}_C)\ge   \mx{q}^H \mx{A}_C \mx{C}_C(r) \mx{C}_C^H(r)\mx{A}_C^H \mx{q} ,
\end{align}
we may replace the first line of \eqref{eq:condition2} as
$$\mx{Q}_C-\mx{C}^{H}_C(r_{\max,C})\mx{A}_C^H \mx{q} \mx{q}^H \mx{A}_C\mx{C}_C(r_{\max,C})\succeq \mx{0}$$

 which due to Shcur complement lemma can be written as
 \begin{align}
   \begin{bmatrix}
\mx{Q}_C&  \mx{C}^{H}_C(r_{\max,C})\mx{A}_C^H \mx{q}\\
\mx{q}^H \mx{A}_C\mx{C}_C(r_{\max,C})&1
\end{bmatrix}\succeq \mx{0}.
 \end{align}
This is due to the fact that the maximum singular value $\sigma_{\max}(\mx{C}_C(r))$ and indeed the maximum eigenvalue of $\mx{C}_C(r) \mx{C}_C^H(r)$ occurs at $r=r_{\max,C}$.
\bibliographystyle{IEEEtran}
\bibliography{references}

\end{document}